\documentclass[twocolumn,pre,aps,10pt]{revtex4-1}

\usepackage{amsfonts,amssymb}
\usepackage[sumlimits,intlimits]{amsmath}
\usepackage{graphics}
\usepackage{graphicx}
\usepackage{epsfig}
\usepackage{mathrsfs}
\usepackage{verbatim}
\usepackage{hyperref}    
\usepackage{bm}          

\newcommand{\be}{\begin{equation}}
\newcommand{\ee}{\end{equation}}

\begin{document}

\title{Theory of volumetric capacitance of an electric double layer supercapacitor}

\date{\today}

\author{Brian Skinner}
\author{Tianran Chen}
\author{M. S. Loth}
\author{B. I. Shklovskii}
\affiliation{Fine Theoretical Physics Institute, University of Minnesota, Minneapolis, Minnesota 55455}

\begin{abstract}

Electric double layer supercapacitors are a fast-rising class of high-power energy storage devices based on porous electrodes immersed in a concentrated electrolyte or ionic liquid. As yet there is no microscopic theory to describe their surprisingly large capacitance per unit volume (volumetric capacitance) of $\sim 100$ F/cm$^3$, nor is there a good understanding of the fundamental limits on volumetric capacitance.  In this paper we present a non-mean-field theory of the volumetric capacitance of a supercapacitor that captures the discrete nature of the ions and the exponential screening of their repulsive interaction by the electrode.  We consider analytically and via Monte-Carlo simulations the case of an electrode made from a good metal and show that in this case the volumetric capacitance can reach the record values.  We also study how the capacitance is reduced when
the electrode is an imperfect metal characterized by some finite
screening radius. Finally, we argue that a carbon electrode,
despite its relatively large linear screening radius, can be approximated as a perfect metal because of its strong nonlinear screening.  In this way the experimentally-measured capacitance values of $\sim 100$ F/cm$^3$ may be understood.

\end{abstract}
\maketitle

\section{Introduction} \label{sec:intro}

The present energy crisis has created a growing demand for
efficient, portable, and high-power energy storage devices.
Electric double layer (EDL) supercapacitors are fast emerging as a
promising potential solution to this problem
\cite{Schindall2007cou}.  In an EDL supercapacitor, energy is
stored at the interface between an electron-conducting (metallic)
electrode and an electrolyte or ionic liquid via the reversible
adsorption of ions onto the electrode surface.  In this way,
counterions adsorbed onto the charged electrode effectively
comprise the second half of a parallel-plane capacitor whose
thickness is equal to the radius $a/2$ of the ions.  If the charge
of these ions is described as a uniformly charged plane, as in the
mean-field approach, then one arrives at a capacitance $C$ which
is equal to \be C_H = 2 \varepsilon_0 \varepsilon A/a,
\label{eq:CH} \ee a result first envisioned by Helmholtz in 1853
\cite{Helmholtz1853ueg, *Helmholtz2004slc}.  Here, $\varepsilon_0$ is
the vacuum permittivity, $\varepsilon$ is the dielectric constant
of the ionic solution, and $A$ is the total surface area of the
electrode.  In mean-field theories of the EDL, $C_H/A$ plays the
role of a maximum possible capacitance per unit area.  As an example, for $\varepsilon = 2$ and $a = 1$ nm Eq.\ (\ref{eq:CH}) gives $C_H/A \approx 3$ $\mu$F/cm$^2$.

\begin{figure}[t!]
\centering
\includegraphics[width=0.35 \textwidth]{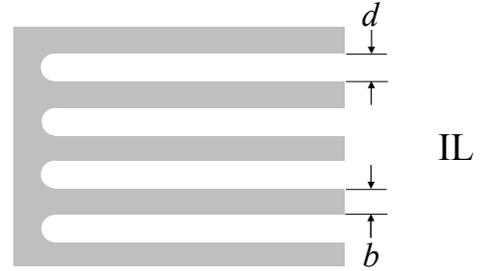}
\caption{A schematic picture of the cross-section of a highly porous supercapacitor electrode.  The solid, metallic electrode (gray area) has deep,
planar pores of width $d$, separated by walls of thickness $b$.
The electrode is open on one side to an ionic liquid (IL) or a
concentrated electrolyte.} \label{fig:comb}
\end{figure}

For practical applications, a supercapacitor is best characterized
not by its capacitance per unit area, $C/A$, but by its
capacitance per unit mass or per unit volume (``volumetric
capacitance").  For this reason, there has been much emphasis on
the development of conducting materials with very high specific
surface area that can be used as electrodes.  Among the more
promising candidates are highly porous carbons \cite{Simon2008mec,
Simon2010csm, Wang2006eoh} and carbon nanotube ``forests"
\cite{Signorelli2009edc}.  In such devices the specific surface
area $\mathbb{S}$ can be as high as $\mathbb{S} = 1000$
m$^2$/cm$^3$.  To understand how this is possible, one can imagine
an electrode with slit-like pores of width $d = 1$ nm separated by
conducting walls with thickness $b = 1$ nm (Fig.\ \ref{fig:comb}).
For the sake of argument, we take this electrode to be the anode; one can
imagine that the cathode is its mirror reflection to the right. In
a supercapacitor device, the anode and cathode are electrically
isolated by a membrane that is penetrable to the ions, so that the well-conducting ionic liquid between them forms EDLs on the tortuous surfaces of both electrodes.  In this way the supercapacitor consists of two double layer
capacitors in series; in this paper we are concerned with
calculating the anode capacitance.

If the electrode in Fig.\ \ref{fig:comb} is placed in contact with
an ionic solution with $\varepsilon = 2$ and $a = 1$ nm, then the
Helmholtz expression of Eq.\ (\ref{eq:CH}) predicts a volumetric
capacitance $\mathbb{C} \approx 30$ F/cm$^3$.  In fact,
capacitance values as large as $\mathbb{C} = 100$ F/cm$^3$ have
been reported for such devices \cite{Simon2008mec,
Largeot2008rbi}.  How is this possible?

In order to resolve this puzzle, let us briefly return to the
problem of a planar, nonporous double layer capacitor.  It has
been shown recently \cite{Loth2010alc} that the
capacitance per unit area of an EDL is not necessarily limited by
the Helmholtz value.  When the charge on a planar electrode is
small enough that adsorbed ions are separated from each other by a
distance much larger than their diameter $a$, the mean-field
approach fails and the effects of electronic polarization of the
electrode surface must be taken into account.  In particular, when
the electrode is made from a good metal, each ion forms an image
charge in the electrode surface.  The ion and its image charge
together make an electric dipole which repels adjacent ions by a
screened $1/r^3$ interaction rather than the normal $1/r$
interaction.  Such a reduced interaction, along with the
positional correlations between adsorbed ions, allows the
capacitance of a single interface to be as much as three times
larger than $C_H$ in practical situations.  The crucial importance of image forces for the structure and capacitance of the EDL has been recognized by a number of previous authors (see, for example, Refs.\ \onlinecite{Torrie1982edl, Bhuiyan2007mpt, Alawneh2008eod} and the very recent publication of Ref.\ \onlinecite{Outhwaite2011ioe}).

\begin{figure}[t!]
\centering
\includegraphics[width=0.45 \textwidth]{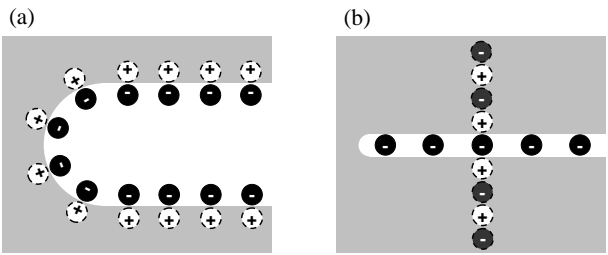}
\caption{Schematic depiction of anions (black circles)
neutralizing a positively charged nanopore in a metallic anode
(solid gray area).  (a) When the pore width is large compared to
the ion diameter, opposite walls of the pore have independent
EDLs and every anion can be said to have a single positive image
charge (white circles with dashed outline).  (b) When the pore
width is comparable to the anion radius, anions form a two-dimensional (2D) charged layer within the pore.  Each anion has an infinite series of image charges (white and dark gray circles with dashed outlines), which produce an exponential interaction between neighboring anions. For clarity of illustration, image charges are shown for one anion
only.} \label{fig:pores}
\end{figure}

The notion of a double-layer comprised of ion-image dipoles is also relevant for describing porous, metallic supercapacitor electrodes (Fig.\ \ref{fig:comb}), provided that the width $d$ of the pores is much larger than the ion diameter $a$, so that opposite walls of a pore have independent, non-interacting EDLs [see Fig.\ \ref{fig:pores}(a)].  For electrodes with such wide pores, enhanced capacitance can be explained using the theory of Ref.\ \onlinecite{Loth2010alc}.  However, in supercapacitors where $d$ is comparable to $a$, EDLs on opposite walls of a nanopore merge and new physics should emerge.  Indeed, recent experiments by Gogotsi and co-workers have demonstrated a surprising increase in the capacitance as the width of pores in a carbon-based electrode is made comparable to the diameter of bare ions in an organic electrolyte \cite{Chmiola2006aii} or in an ionic liquid \cite{Largeot2008rbi}.

In order to explain these results, one can try to extend the
Helmholtz mean-field approach to the case of a narrow pore by
replacing the charge of ions in the pore by two identical, coinciding, uniformly-charged planes located midway between the pore's two walls (see the result of a similar approach for cylindrical pores in Ref.\ \onlinecite{Huang2008umn}).  In this picture, each of the charged planes forms a Helmholtz capacitor with one of the pore's metal walls, so that the total capacitance is $2 C_H$, as it would be for a much wider pore.  Thus, the mean-field approach cannot capture the unique effect of narrow pores.

Going beyond the mean-field level, however, one can recall that
charges confined within a narrow nanopore create an infinite
series of image charges in the two conducting electrode walls
[Fig.\ \ref{fig:pores}(b)].  This leads to an interaction between ions which decays exponentially with the distance between them \footnote{
This exponential interaction has been used previously
\cite{Koulakov1998csa} in order to describe mesoscopic
oscillations in the capacitance of a disc-like island of a
classical two-dimensional electron gas between two metallic,
planar electrodes as a function of the number of electrons in the
island.  Such an electron gas, confined at the periphery by a
cylindrical later gate, constitutes a vertical quantum dot
\cite{Oosterkamp1999mda}. }.

In a recent paper, Kondrat and Kornyshev \cite{Kondrat2011ssi}
recognized that such an exponentially suppressed interaction can
lead to large capacitance $C > C_H$ in the nanopore for reasons
similar to those discussed in Ref.\ \onlinecite{
Loth2010alc}.  However, in calculating the capacitance, the
authors of Ref.\ \onlinecite{Kondrat2011ssi} imagined that the pore is filled with many anions and cations and they used a description where the small net charge is spread equally among \emph{all} ions in the pore. The total electrostatic energy was then calculated using the exponential interaction evaluated at the average distance between all ions.  Such an approach can be called a semi-mean-field approximation and does not lead to quantitatively correct results, as we will show below.

In the present paper, as in Ref.\ \onlinecite{
Loth2010alc}, we completely abandon the mean-field approach in
order to address a fundamental question: how large can the
volumetric capacitance of an EDL supercapacitor be?  We construct
a theory which takes into account correlations between discrete
anions and the screening of the Coulomb interaction by the conducting electrode surface.  We consider explicitly two cases for the electrode material.  First, we examine the case where the
electrode can be considered a perfect metal (Sec.\
\ref{sec:metal}) and we verify our theoretical predictions with a
simple Monte Carlo (MC) simulation (Sec.\ \ref{sec:MC}).  We find
that under realistic circumstances the capacitance per unit area
can be up to ten times the Helmholtz value, with the corresponding
volumetric capacitance as large as $150$ F/cm$^2$.  Secondly, we
analyze the case where the electrode is an imperfect metal with a
screening radius comparable to the pore width (Sec.\
\ref{sec:3Dscreening}).  The crossover between this theory and
that of Sec.\ \ref{sec:metal} is carefully discussed.  In
Sec.\ \ref{sec:graphite} we consider how our theory applies to
supercapacitors made with graphitic carbon electrodes, and we
argue that at not-too-small voltages the graphite is well
approximated as a good metal due to the strong effect of nonlinear
screening.  We close in Sec.\ \ref{sec:balls} by briefly examining a different model of a porous supercapacitor, where the electrode is made from a random assembly of conducting spheres that are three-dimensionally connected, and show that very large volumetric capacitance can result in this situation as well.

\section{Capacitance of a single 2D metal pore} \label{sec:metal}

In this section we consider an electrode made from a perfect metal
which has deep, planar pores of width $d$ (Fig.\ \ref{fig:comb}). Such pores are assumed
to be in contact with an ionic liquid described by the restricted
primitive model: a neutral mixture of hard-core monovalent ions
with the same diameter $a$.  We assume that $a \leq d$, so that
ions can enter the pores.  We also assume that $d-a \ll a$, so that
ions in the pore can be described as a 2D liquid.  A voltage source
provides the positive potential difference $V$ between the
electrode and the bulk of the ionic liquid that attracts anions into
the pore.  If $Q$ is the amount of electronic charge that has
moved through the voltage source onto the electrode relative to the state at $V = 0$, then the differential capacitance of the EDL
is defined as $C = dQ/dV$.

In principle, at $V = 0$ the pore may already contain some finite and equal number of anions and cations.  In this case, a reliable analytical calculation of the total electrostatic energy $U(Q)$, which is necessary for calculating the capacitance, is very difficult.  Therefore, the effect of allowing both ionic species to simultaneously enter the pore is examined only numerically at the end of Sec.\ \ref{sec:MC}.

This paper concentrates instead on the case when the pore
is empty at $V = 0$.  This situation results when the chemical potential of ions in the ionic liquid is lower than the free energy per ion in a filled, neutral pore.  Such a difference in chemical potential can arise from two sources.  First, ions in a three-dimensional (3D) ionic liquid are surrounded by a larger number of oppositely-charged neighbors, which lowers the interaction part of the chemical potential outside the pore.  Second, when the width $d$ of the pore is close to the ion diameter $a$, the entropic contribution to the chemical potential inside the pore increases sharply.  For the case of electrolyte solutions, there is also a positive contribution to the chemical potential associated with the necessity of stripping the solvation shell from each ion that enters the pore.

We therefore assume that the pores are empty at $V = 0$.  As the voltage is increased from zero, the pores remain empty until some finite voltage $V = V_t$.  At $V > V_t$ the pores of the anode begin to fill with anions, while cations remain away from the anode.  This picture allows us to formulate a simple analytical calculation of the total energy $U$, presented below, based on the repulsion between anions in the pore.

Our general approach to calculating the capacitance is as follows.
We first describe the total electrostatic energy $U(n)$ associated
with the lowest energy configuration of $n$ anions per unit area
in the pore.  If entropic effects are ignored, then the value of
the charge $Q$ of the pore is that which minimizes the system's
total energy $U - QV$, where the term $-QV$ represents the work
done by the voltage source.
Using the equilibrium condition $d(U - QV)/dQ = 0$ along with $Q =
e A n$ gives 
\be 
V = \frac{d U}{d Q} = \frac{1}{eA} \frac{d U}{dn}. 
\label{eq:Vdef} 
\ee 
The differential capacitance of the
pore $C = (dV/dQ)^{-1}$ can therefore be written 
\be 
C = e^2 A^2 \left( \frac{d^2 U}{d n^2} \right)^{-1}. 
\label{eq:Cdef} 
\ee 
The capacitance can be expressed as a function of voltage, $C(V)$, by
combining Eqs.\ (\ref{eq:Vdef}) and (\ref{eq:Cdef}).

In the remainder of this section we first calculate the
capacitance of the pore in the zero temperature limit and then
estimate the effect of the ions' finite thermal energy.

We begin our theoretical description by noting that a point charge
$e$ located in the plane halfway between two metal walls creates
an electric potential within that plane equal to
\cite{Smythe1939sad} 
\be 
\phi(r) = \frac{e}{\pi \varepsilon_0 \varepsilon d} \sum_{n = 1}^{\infty} K_0 [ \pi (2n - 1) r/d ].
\label{eq:phiexact} 
\ee 
Here, $r$ is the radial distance from the
point charge, $d$ is the distance between the metal walls (the
pore width), and $K_0(x)$ is the zeroth order modified Bessel
function of the second kind.  At distances $r > d$, Eq.\
(\ref{eq:phiexact}) can be expanded to lowest order to give 
\be
\phi(r) \simeq \frac{2 \sqrt{2} \exp[-\pi r/d]}{\sqrt{r/d}}
\frac{e}{4 \pi \varepsilon_0 \varepsilon d}. 
\label{eq:phi} 
\ee
Since the sub-leading-order term of Eq.\ (\ref{eq:phiexact}) is
exponentially smaller than that of Eq.\ (\ref{eq:phi}), this
approximation has a negligible effect on the capacitance and we
use Eq.\ (\ref{eq:phi}) everywhere in further calculations.

When a given area density $n$ of anions is inside the metal pore,
the repulsive interaction between anions induces strong positional
correlations.  In their lowest energy configuration, the anions
form a strongly-correlated liquid, reminiscent of a 2D Wigner
crystal, where anions are separated from their nearest neighbors
by a well-defined spacing $\sim n^{-1/2}$.  In such an arrangement
the total repulsive energy among anions is minimized while
maintaining the area density required to neutralize the electrode.

If we postulate a crystalline arrangement of the anions, then
the electrostatic energy $U$ of this state can be calculated
exactly by making use of the interaction potential in Eq.\
(\ref{eq:phi}).  Due to the short-ranged nature of the
interaction, this energy is well approximated by considering only
nearest-neighbor interactions in a square lattice of anions. Such
an approach gives
\begin{eqnarray}
U(n) & = & 2 n A e \phi(n^{-1/2}) - (\mu-u) n A \nonumber \\
 & = & 4 \sqrt{2} \frac{A}{d^2} (n d^2)^{5/4} \exp\left[-\pi/\sqrt{nd^2}\right] \frac{e^2}{4 \pi \varepsilon_0 \varepsilon d} 
\label{eq:Upore} \\
& & - (\mu-u) n A. \nonumber
\end{eqnarray}
The term $-(\mu-u) n A$ takes into account the voltage-independent energy associated with bringing each anion from the bulk of the ionic liquid into the pore; $\mu$ is the chemical potential of ions in the bulk of the ion liquid and $u$ is the self-energy of an anion in the pore.  The term $-(\mu-u) n A$ is linear in $n$ and therefore, by Eq.\ (\ref{eq:Cdef}), disappears from the capacitance. Its only effect is to produce a finite threshold voltage $V_t= -(\mu - u)/e$ required to bring anions into the metal pore, as discussed above.  Our theory treatment assumes that $-(\mu - u) > 0$.

Taking the derivative $dU/dQ$ as in Eq.\ (\ref{eq:Vdef}), we find
an expression for the voltage in terms of the ion density: 
\be 
V - V_t \simeq \frac{2 \pi \sqrt{2} \exp[-\pi/\sqrt{n d^2}]}{(n
d^2)^{1/4}} \frac{e}{4 \pi \varepsilon_0 \varepsilon d}.
\label{eq:Vpore} 
\ee 
Similarly, the capacitance can be evaluated
by Eq.\ (\ref{eq:Cdef}), which gives 
\be 
C \simeq \frac{\sqrt{2}a}{\pi d } (n d^2)^{7/4} \exp[\pi/\sqrt{n d^2}] C_H.
\label{eq:Cpore} 
\ee 
In the limit $n \ll 1/d^2$, Eqs.\ (\ref{eq:Vpore}) and (\ref{eq:Cpore}) can be combined to give an analytical expression for the capacitance as a function of voltage at small $V - V_t$: 
\be 
C  \simeq 32 \pi^3 \left( \frac{e/4 \pi \varepsilon_0 \varepsilon d}{V - V_t} \right)  \ln^{-3} \left[ 8 \pi^2 \left( \frac{e/4 \pi \varepsilon_0 \varepsilon d}{V - V_t} \right)^{2}
\right] \frac{a}{d} C_H. 
\label{eq:C-smallV-pore} 
\ee 
For larger voltages corresponding to $(V - V_t) / (e/4\pi\varepsilon_0\varepsilon d) \gtrsim 0.1$ the capacitance is well-described by the power law relation 
\be 
C \simeq 3.5 \left( \frac{V - V_t}{e/4\pi\varepsilon_0\varepsilon d}
\right)^{-0.4} \frac{a}{d} C_H. 
\label{eq:CVpore} 
\ee

\begin{figure}[htb]
\centering
\includegraphics[width=0.45 \textwidth]{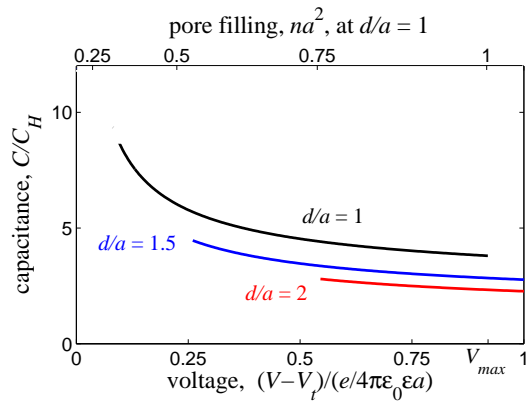}
\caption{(Color online). The capacitance in units of $C_H$ of a 2D metal nanopore with width $d$, plotted as a function of
dimensionless voltage  (lower axis) according to Eqs.\ (\ref{eq:Vpore}) and (\ref{eq:Cpore}).  Different curves are labeled by their corresponding value of $d/a$ and are truncated at their corresponding value of $C_{max}$ as given by Eq.\ (\ref{eq:Cmax}); here, $T^* a/d = 0.04$.  $V_{max}$ is shown for $d/a = 1$, and the top axis shows the ion concentration for this same case.
} \label{fig:C-V}
\end{figure}

Eqs.\ (\ref{eq:Cpore}) and (\ref{eq:C-smallV-pore}) suggest that
at low ion density (or small $V - V_t$) the capacitance can be
much larger than the Helmholtz value.  This result can be understood physically by noting that at such low ion densities the fractional coverage of excess ions on the electrode surface $n a^2 \ll 1$, so that it is incorrect to think of the EDL in the mean-field way: as a uniform layer of surface charge.
Rather, the neutralizing ionic charge consists of discrete ions
whose interaction is exponentially small due to the aggressive
screening by the metal pore.  Positional correlations among these
ions help them to avoid each other, resulting in a lower energy than what is possible in mean-field descriptions of the EDL and therefore in larger capacitance that is not limited by the physical distance $a/2$ between the electrode and its countercharge.  With growing ion density, the capacitance decreases, until at some finite voltage $V_{max}$ thedensity of ions in the pore reaches its steric limit: $n \simeq 1/a^2$.  By Eq.\ (\ref{eq:Vpore}), 
\be 
V_{max} \simeq 2\pi \sqrt{\frac{2a}{d}} \exp[-\pi a/d] \frac{e}{4\pi \varepsilon_0 \varepsilon d}. 
\label{eq:Vmax} 
\ee 
Fig.\ \ref{fig:C-V} shows the capacitance as a function of voltage, $C(V)$, plotted for the cases $d = a, 1.5 a, 2a$.

If the width of pores in the electrode is increased, the
capacitance decreases, as shown in Fig.\ \ref{fig:C-V}.  In the
limit where the pore thickness $d \gg a$, as in Fig.\
\ref{fig:pores}(a), the capacitance can be described using a
theory of independent EDLs comprised of ion-image dipoles.  Such
an approach gives $C \approx 1.3 C_H$ per interface
\cite{Loth2010alc} at $n a^2 = 1$ (the relatively flat tail of
the $C$--$V$ curve) and $C_{max} \approx 3 C_H$ per interface at $n \rightarrow 0$, so that the total capacitance per pore is smaller than the result shown in Fig.\ \ref{fig:C-V} by more than two times.  As mentioned above, this ``anomalous" increase in the capacitance for narrow pores is the result of the strong, exponential screening that results from the presence of two close metal walls [Fig.\ \ref{fig:pores}(b)].

Formally, Eq.\ (\ref{eq:Cpore}) diverges as the density of ions
vanishes ($V - V_t$ goes to zero).  Of course, this expression
neglects entropic effects among the ions, which are important in the limit where ions in the pore are so sparse that their typical interaction energy is smaller than the thermal energy $k_BT$.  At such low densities the correlated, lattice-type structure of ions in the pore disappears and we obtain a finite capacitance at $(V - V_t) \rightarrow 0$.

In order to estimate the value of this capacitance maximum, we
note that when $e \phi(n^{-1/2}) \ll k_BT$ the total free energy $F$ can be written using a truncated virial expansion: 
\be 
F \simeq F_{id} + A n^2 k_BT B(T) - e A n (V - V_t). 
\label{eq:Fvirial}
\ee 
Here, $F_{id} = A n k_BT \ln(n a^2)$ is the free energy of a two-dimensional ideal gas and $B(T)$ is the second virial coefficient.  $B(T)$ is calculated from the interaction energy $e
\phi(r)$ between two ions [Eq.\ (\ref{eq:phi})] as
\begin{eqnarray}
B(T) &= & \frac12 \int_0^{\infty} \left(1 - \exp\left[-\frac{e \phi(r)}{k_BT}\right] \right) 2 \pi r dr \\
& \simeq & \frac{d^2}{8 \pi} \ln^2 \left[ \frac{16 \pi}{(T^*)^2} \right].
\label{eq:B}
\end{eqnarray}
Here, $T^*$ is defined as the dimensionless temperature
\be
T^* = \frac{k_BT}{e^2/4\pi \varepsilon_0 \varepsilon d}.
\label{eq:Tstar}
\ee
As an example, room temperature corresponds to $T^* \approx 0.04$ for an ionic solution with dielectric constant $\varepsilon = 2$ in a pore with $d = 1$ nm.  At temperatures that are not very large, $T^* \lesssim 0.5$, the virial coefficient $B(T)$ is larger than the physical area $\pi a^2/4$ occupied by each ion, so that the hard-core interaction between ions is unimportant for the virial expansion.

As before, we can use the equilibrium condition $\partial F/\partial n = 0$ to give a relation between the voltage and the ion density $n$:
\be 
e (V - V_t) = k_BT \left[ 2nB(T) - \ln\left( 1/na^2 \right) \right].
\label{eq:Vvirial}
\ee 
The capacitance can also be related to $n$ according to $C = e^2 A^2 (\partial^2 F/\partial n^2)^{-1}$, which gives
\be 
C = \frac{n d^2}{T^*[1 + 2 n B(T)]} \frac{a}{d} C_H.
\label{eq:Cvirialn}
\ee 
According to Eq.\ (\ref{eq:Vvirial}), in the limit $(V - V_t) = 0$ the ion density approaches $n = W_0[2 B(T)/a^2]/2 B(T)$, where $W_0[x]$ is the principle branch of the Lambert $W$ function ($W_0[x] \approx \ln x$ for $x \gg 1$).  Over the experimentally relevant range of temperature $0.03 < T^* < 1$, the value of $W_0[2 B(T)/a^2] \approx 1$, so that Eqs.\ (\ref{eq:Vvirial}) and (\ref{eq:Cvirialn}) can be combined to give the following approximate relation for the capacitance at $V = V_t$ as a function of temperature:
\be 
C_{max}(T) = \frac{\gamma}{T^* \ln^2 \left[ 16 \pi / (T^*)^2\right]} \frac{a}{d} C_H. 
\label{eq:Cmax} 
\ee
Here $\gamma$ is a numerical constant; Eq.\ (\ref{eq:B}) suggests $\gamma = 4 \pi^2 \approx 39$, while MC simulations (see Fig.\ \ref{fig:MC-compare} in the following section) give $\gamma = 34 \pm 1$.  This is a surprisingly good agreement, considering that $C_{max}$ is determined by a relatively large ion density $n \sim 1/B(T)$, which is at the limit of applicability of the truncated virial expansion of Eq.\ (\ref{eq:Fvirial}).  Indeed, Eq.\ (\ref{eq:Fvirial}) is applicable only in the ``gas phase" corresponding to $n B(T) \ll 1$, which is realized at $V < V_t$.  On the other hand, Eq.\ (\ref{eq:Fvirial}) fails completely in the high-density correlated liquid phase, where $n B(T) \gg 1$ and the potential energy of repulsion between anions [see Eq.\ (\ref{eq:Upore})] dominates the entropic contribution to the free energy.  A more complete theory of the capacitance at finite temperature and large ion density would require a theory of the free energy of ions in the liquid state and cannot be captured by the truncated virial expansion presented here.  The zero temperature analytical result of Eq.\ (\ref{eq:CVpore}), however, should be accurate in the limit where the thermal energy $k_BT$ is small compared to the typical interaction energy $e \phi(n^{-1/2})$.  At room temperature and for $d = 1$ nm and $\varepsilon = 2$, this corresponds to moderately large ion density $nd^2 \gtrsim 0.3$.

The above results for the capacitance of a single metal pore can
be used to calculate the total capacitance of the electrode by
multiplying by the number of pores in the electrode.  Thus,
the total volumetric capacitance $\mathbb{C}$ is given by $\mathbb{C} = (\varepsilon_0 \varepsilon \mathbb{S}/a) \cdot C/C_H$.  As an example we can consider the electrode depicted in Fig.\ \ref{fig:comb}, with $d = b = 1$ nm, in contact with an ionic liquid with $\varepsilon = 2$
and $a = d$.  For such a capacitor the mostly flat tail of the
$C$--$V$ curve of Fig.\ \ref{fig:C-V}, where $C/C_H \approx 4.5$,
corresponds to $\mathbb{C} \approx 90$ F/cm$^3$.  If the
electrode can be treated as a perfect metal, then such a capacitor
would demonstrate a peak in the differential capacitance at a
particular voltage $V_t$, as in Fig.\ \ref{fig:C-V}.  For $T^* =
0.04$, $\mathbb{C}_{max} \approx 150$ F/cm$^3$.

\section{Monte Carlo simulation of a 2D metal pore} \label{sec:MC}

In order to verify the theoretical predictions of the previous
section, we perform MC simulations of a 2D metal pore open to a
reservoir of positive and negative hard-sphere ions.  Our general
approach is to use the Grand Canonical Monte Carlo (GCMC) method
to impose a difference in chemical potential between positive and
negative ions in the system, thereby simulating an applied voltage
$V$.  Specifically, the chemical potential of each ion type is
specified according to 
\be 
\mu_\pm = \mu \mp e V,
\label{eq:chem_pot} 
\ee where $\mu$ is the chemical potential of the reservoir.  The number of each ion species is allowed to fluctuate with time.  We use our simulation to measure the resulting equilibrium number of positive and negative charges in the pore at a given $V$, which defines the net charge $Q(V)$.  The capacitance of the pore is calculated by the discrete derivative $dQ/dV$.

The details of our simulation method are as follows.  We begin
each simulation by randomly placing 100 of each type of ion on a
square 2D plane of area $A = 20 \times 20 d^2$ and stipulating the
dimensionless temperature $T^*$ [see Eq.\ (\ref{eq:Tstar})] and
the dimensionless voltage $V^* = V/(e/4 \pi \varepsilon_0
\varepsilon d)$.  The ion diameter $a$ is taken to be equal to the
pore width $d$.  Before any data is taken, ions are allowed to
take $10^5$ GCMC steps to reach equilibrium.  The number of
positive and negative ions, $M_+$ and $M_-$, respectively, are
then averaged over the following $30$ to $50 \times 10^6$ GCMC
steps. The charge of the electrode is defined as $Q = e(M_- -
M_+)$ and the capacitance is given by the discrete derivative
$C(V) \approx [Q(V + \Delta V/2) - Q(V - \Delta V/2)]/\Delta V$.
Care is taken to ensure that all results are independent of the
initial ion configuration.

Following the standard GCMC procedure \cite{Frenkel2001ums}, one
GCMC step consists of either an attempted move by a
randomly-chosen ion or the attempted addition or removal of an ion
from the system.  Attempted moves occur more often than attempted
addition/removal at a ratio of $9:1$.  We give the simulation area periodic boundaries, so that an ion leaving one edge enters
at the opposite edge.  The total electrostatic energy $U_{tot}$ of
a given configuration of ions is calculated as 
\be 
U_{tot} = \frac{1}{2}\sum_{i,j}q_i q_j e \phi(r_{i,j}), \label{eq:Utotal}
\ee 
where $r_{ij}$ is the distance between ions $i$ and $j$ (found
using the minimum image convention \cite{Valleau1980pme}), $q_i =
\pm 1$ is the sign of ion $i$, and $\phi(r)$ is the interaction
law given by Eq.\ (\ref{eq:phi}).  Attempted moves and
addition/removal events are accepted and rejected based on the
corresponding change in $U_{tot}$, as given by the traditional
acceptance rules for GCMC \cite{Frenkel2001ums}.  The ions are
treated as hard spheres, so that only those
moves/additions/removals resulting in non-overlapping ions are
accepted.  For a more detailed discussion of the GCMC method see
Refs. \onlinecite{Valleau1980pme, Valleau1980pmea}.

\begin{figure}[htb]
\centering
\includegraphics[width=0.45 \textwidth]{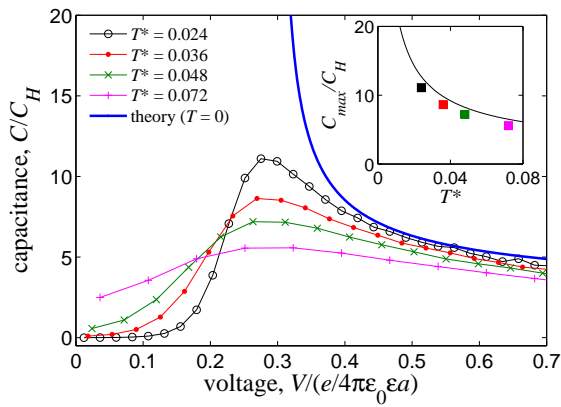}
\caption{(Color online). MC results for the capacitance $C$, in units of $C_H$, of a 2D metal pore with $V_t^* = 0.3$, plotted as a function of voltage for a range of temperatures  $T^*$.  The thick line shows the prediction of our analytical theory. For each temperature, the capacitance attains its maximum at $V^* \approx V_t^* = 0.3$.  These maximum capacitances (shown in the inset by filled squares) can be described by Eq.\ (\ref{eq:Cmax}) with $\gamma = 34 \pm 1$ (the theoretical curve with $\gamma = 4 \pi^2$ is shown by the thin line). } \label{fig:MC-compare}
\end{figure}

Fig.\ \ref{fig:MC-compare} shows the capacitance measured by our
GCMC simulation as a function of voltage for a system with
$V_t^* = V_t/(e/4 \pi \varepsilon_0 \varepsilon d) = -(\mu^*-u^*)
= 0.3$, calculated at four different dimensionless temperatures.
In this situation, the pore is essentially empty of ions at $V =
0$ and at positive voltages contains only one ionic species, so
that its capacitance is well described by the analytical theory of
Sec.\ \ref{sec:metal}.  Both the voltage and temperature
dependence of the capacitance correspond closely to analytical
predictions. Larger voltages could not be examined by our
simulation since these correspond to large ion fillings $na^2 \sim
1$ (see the top axis of Fig.\ \ref{fig:C-V}), at which the
simulation fails to reach equilibrium in a reasonable amount of
time.

So far we have dealt only with pores that are empty at $V = 0$.  In the remainder of this section we use the GCMC simulation to qualitatively examine a pore containing a substantial, neutral concentration of both anions and cations at $V = 0$.  To arrive at such a situation, one should increase the chemical potential $\mu$, causing the value of $V_t$ to decline
\footnote{
In our MC simulations, we observe the value of the threshold voltage $V_t^*$ to be equal to $-(\mu^* - u^*)$ when $\mu^*$ is large negative.  At some particular value of $-(\mu^* - u^*) > 0$, however, the threshold voltage abruptly disappears and the pore is spontaneously filled with ions at zero voltage.  Such a transition can be understood qualitatively by considering that when $(\mu^* - u^*) < 0$ it is not energetically favorable for single ions to enter the pore, but ions may still enter the pore in neutral pairs or larger neutral clusters.  Thus, the location of the $V = 0$ empty-to-filled pore transition is related to the electrostatic energy per particle of an ion in a neutral cluster.  For $T^*$ between $0.02$ and $0.07$, we observe the threshold voltage to disappear at $-(\mu^* - u^*)$ between $0.08$ and $0.12$. }.

\begin{figure}[htb]
\centering
\includegraphics[width=0.45 \textwidth]{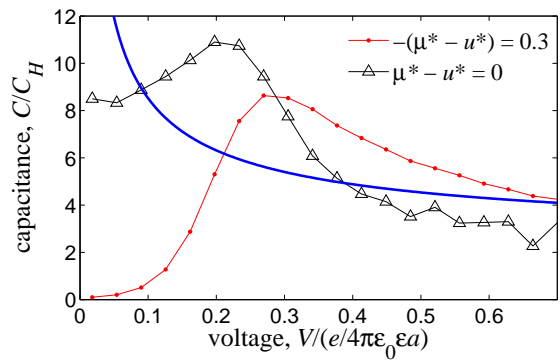}
\caption{(Color online). MC results for the capacitance of a 2D metal pore with $T^* = 0.04$ at $\mu^*-u^*
= 0$ (triangles) and $-(\mu^*-u^*)= 0.3$ (circles). The thick line is the prediction of Eq.\ (\ref{eq:CVpore}) with $V_t^* = -(\mu^*-u^*) = 0$. } \label{fig:different_mu}
\end{figure}

Fig.\ \ref{fig:different_mu} shows, as an example, the capacitance of a pore with $\mu^*-u^* = 0$, which corresponds to moderately large ion filling at $V = 0$.  We also show a system with $V_t^* = 0.3$ for comparison.  The pore with $\mu^* - u^* = 0$ (triangles) is more than half-filled at zero voltage: $(M_+ + M_-)a^2/L^2 \approx 0.63$.  As the voltage is increased from zero, cations are driven out of the pore and anions are attracted to the pore until at $V^* \gtrsim 0.35$ only anions remain in the pore and the capacitance is reasonably well described by our analytic treatment of the previous section (as shown by the solid line).  At $V^* < 0.35$, on the other hand, the strong attraction between cations and anions affects the capacitance.  As a rough approach to explaining this data, one may imagine that at small voltage the net ionic charge consists of a small number $Q/e$ of ``excess anions" on the background of a large number of neutral, tightly-bound cation-anion pairs.  These excess anions seek to maximize their distance from each other by forming a correlated, Wigner crystal-like arrangement in a way that is similar to the description of the previous section.  Under this description, one may expect the same analytical theory to hold as for large negative $\mu^*$, since the neutral pairs are essentially non-interacting and therefore play only a small role in determining the capacitance.  Fig.\ \ref{fig:different_mu} suggests that this approach gives a reasonably accurate description of the finite temperature truncation of the capacitance divergence.  Indeed, the capacitance in limit $V^* + (\mu^* - u^*) = 0$ is very similar for the two $C$--$V$ curves.  However, this approach does not explain the weak capacitance maximum at $V^* \approx 0.2$ in the $C$--$V$ curve corresponding to $\mu^* - u^* = 0$, which remains a puzzle.

This qualitative explanation of the filled pore data is similar to that of Ref.\ \onlinecite{Loth2010alc}, but is fundamentally different from the approach of Ref.\ \onlinecite{Kondrat2011ssi}.  These authors assumed that the charge of excess anions is spread equally over all ions in the pore, with each ion getting a small fraction $\delta e$ of the electron charge $e$.  They further assumed [see their Eq.\ (3)] that every ion interacts with its nearest neighbors via the exponential interaction $(\phi(r)/e)(\delta e)^2$, where $\phi(r)$ is given by Eq.\ (\ref{eq:phi}). Such a semi-mean-field approximation makes the total repulsive energy of excess anions larger than in our description of Sec.\ \ref{sec:metal}, since the distance between interacting ions is smaller and this changes the exponential factor of $\phi(r)$. Therefore, it seems reasonable that the pore capacitance evaluated in Ref.\ \onlinecite{Kondrat2011ssi} is three times smaller than in our Fig.\ \ref{fig:different_mu}.

\section{Capacitance of a porous imperfect metal} \label{sec:3Dscreening}

Thus far we have calculated the capacitance in situations where the electrode can be considered a perfect metal, or in other words where the electrode has a vanishing electronic screening radius.  In this section we examine what happens to the capacitance when the electrode is not a perfect metal, but instead has a finite screening linear radius $r_s$, given by 
\be 
r_s = \sqrt{ \frac{\varepsilon_0 \varepsilon}{e^2 \nu(\mu_F)} }, 
\label{eq:rs}
\ee 
where $\nu(\mu_F)$ is the electron density of states at the
Fermi energy $\mu_F$ of the electrode.

Below we consider separately two limiting cases for $r_s$: (i) where $r_s \ll b$, the typical thickness of the wall separating adjacent pores, so that adjacent pores can be considered non-interacting, and (ii) where $r_s > b$, so that adsorbed ions interact three-dimensionally.

\begin{figure}[htb]
\centering
\includegraphics[width=0.25 \textwidth]{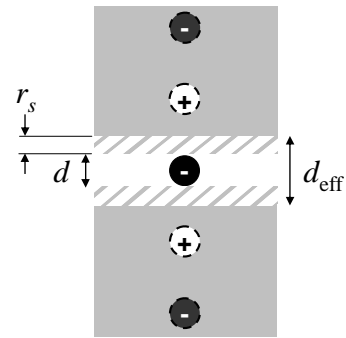}
\caption{A schematic portrayal of the effect of finite screening radius $r_s \ll b$ in a conducting electrode (solid gray area).  The reflection plane for image charges (black and white circles with dashed outlines) of an adsorbed cation (black circle) is shifted by a distance $r_s$ from the wall of the pore (hatched area).  Compare to Fig.\ \ref{fig:pores}(b).} \label{fig:smallrs}
\end{figure}

In the case where $r_s \ll b$, there is no interaction between adjacent pores and the volumetric capacitance can still be calculated as in the previous section, by considering the capacitance of a single pore.  In this case the effect of finite screening radius is to shift the position of the reflection plane for image charges by a distance $r_s$  beyond the surface of the pore wall \cite{Loth2009nsb}.  This reflection plane coincides with the ``electrostatic surface" of the pore, at which the center of gravity of the surface charge is effectively located.  That is, a charge in the center of the pore becomes separated from its image charge by a distance $d + 2 r_s$, as shown in Fig.\ \ref{fig:smallrs}.  In this way the interaction between neighboring ions is stronger than what is given by Eq.\ (\ref{eq:phi}) and the capacitance of the pore is reduced as compared to the results in Sec.\ \ref{sec:metal}.  One can easily calculate the effect this has on the capacitance by replacing $d$ with $d_\text{eff} = d + 2 r_s$ in Eqs.\ (\ref{eq:phiexact}) -- (\ref{eq:Cmax}).  In other words, allowing for finite screening radius $r_s \ll b$ in the electrodes has the same effect as increasing the pore width (which is examined in Fig.\ \ref{fig:C-V}).  For example, a pore with width $d = 1$ nm and $r_s = 0.25$ nm would have $d_\text{eff}/d \approx 1.5$, and the capacitance would correspond to the middle (blue) curve in Fig. \ref{fig:C-V}.

In the opposite limit, $r_s > b$, the wall of the pore does not completely screen the charge of an adsorbed cation and ions in adjacent pores interact with each other.  In this limit ions interact three-dimensionally via a Yukawa-like potential
\be
\phi(r) = \frac{e }{4 \pi \varepsilon_0 \varepsilon r} \exp \left[-\frac{r}{R_s} \right] ,
\label{eq:Yukawa}
\ee
where $R_s$ is the three-dimensional (3D) screening radius, determined from the volume-averaged density of states $\nu(\mu_F) b/(b+d)$:
\be
R_s = \sqrt{\frac{\varepsilon_0 \varepsilon}{e^2 \nu(\mu_F)} \left(1 + \frac{d}{b} \right) } = r_s \sqrt{1 + \frac{d}{b}}.
\ee
For $d = b$ (as in Fig.\ \ref{fig:comb}), we get $R_s = \sqrt{2} r_s$.

In the limit of $R_s \gg N^{-1/3}$, where $N$ is the three-dimensional concentration of ions inside the electrode, the electric potential $\Phi$ is uniform throughout the volume of the electrode and is given by
\be
\Phi = e N \int_0^\infty \phi(r) 4 \pi r^2 dr = \frac{e N R_s^2}{\varepsilon_0 \varepsilon}.
\ee
This gives for the volumetric capacitance
\be
\mathbb{C} = \frac{\varepsilon_0 \varepsilon}{R_s^2}.
\label{eq:C3Dmf}
\ee
This result was first derived as the volumetric capacitance of charged DNA condensates with cationic polyelectrolytes in salty water \cite{Zhang2004pdo, Zhang2005pfo, Skinner2009nso}.

\begin{figure}[htb]
\centering
\includegraphics[width=0.3 \textwidth]{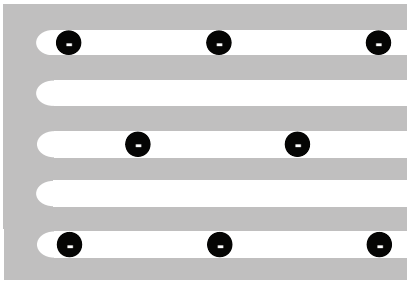}
\caption{ A schematic depiction of anions (black circles) arranging themselves within a porous supercapacitor electrode (solid gray area) in the limit $r_s > b$.  Here, ions interact three-dimensionally, leading to a 3D Wigner-crystal-like arrangement of ions in the ground state. } \label{fig:3D-schematic}
\end{figure}

At smaller $R_s < N^{-1/3}$, the discreteness of the ions plays an important role.  In their lowest energy state, the ions form a correlated, 3D liquid in which they maximize their separation from each other while neutralizing the bulk charge of the electrode, as shown in Fig.\ \ref{fig:3D-schematic}.  As in the previous section, we can calculate the capacitance by postulating a crystalline arrangement of the ions (a 3D Wigner crystal) and calculating the total electrostatic energy $U$.  The capacitance can then be found using the 3D analogue of Eqs.\ (\ref{eq:Vdef}) and (\ref{eq:Cdef}), namely
\begin{eqnarray}
V = \frac{d U}{d Q} = \frac{1}{e\mathbb{V}} \frac{d U}{dN},
\label{eq:Vdef3D} \\
C = e^2 \mathbb{V}^2  \left( \frac{d^2 U}{d N^2} \right)^{-1}, \label{eq:Cdef3D}
\end{eqnarray}
where $\mathbb{V}$ is the electrode volume.

\begin{figure}[htb]
\centering
\includegraphics[width=0.45 \textwidth]{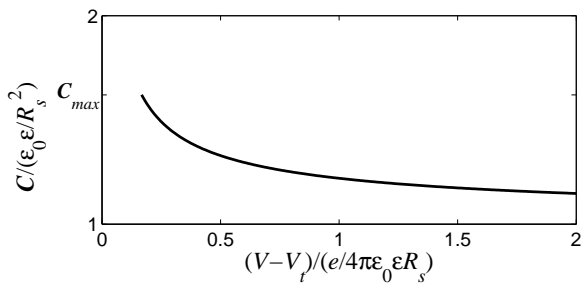}
\caption{The volumetric capacitance $\mathbb{C}$, in units of $\varepsilon_0 \varepsilon/R_s^2$, for the case where ionic charges obey the interaction law of Eq.\ (\ref{eq:Yukawa}), plotted as a function of dimensionless voltage.  The curve is truncated at $\mathbb{C}_{max}$ as given by Eq.\ (\ref{eq:Cmax3D}). } \label{fig:C-V-3D}
\end{figure}

This approach allows one to calculate $\mathbb{C}$ as a function of ion density $N$ and as a function of voltage $V$.  Using a numeric evaluation of the total energy $U$ gives a capacitance that can be accurately fitted to the following power-law form at $(V - V_t)/(e/4\pi\varepsilon_0\varepsilon R_s) > 0.01$:
\be
\mathbb{C}(V) = \left[ 1 + 0.22 \left( \frac{V - V_t}{e/4\pi\varepsilon_0\varepsilon R_s} \right)^{-0.58} \right] \frac{\varepsilon_0 \varepsilon}{R_s^2}.
\label{eq:CV3D}
\ee
This expression is plotted in Fig.\ \ref{fig:C-V-3D}.

Eq.\ (\ref{eq:CV3D}) suggests that the capacitance diverges at small $V - V_t$, as in the case of 2D pores.  To understand why this is the case, we can consider the limit where ions are sufficiently sparse that their separation is much larger than $R_s$.  In this limit only the interactions between nearest-neighbors of the 3D Wigner crystal contribute significantly to the total energy.  If we imagine a cubic lattice arrangement of ions, then we arrive at an energy
\be
U \simeq 3 \mathbb{V} \frac{e^2 N^{4/3} }{4 \pi \varepsilon_0 \varepsilon} \exp \left[ - \frac{1}{N^{1/3} R_s} \right].
\ee
By Eqs.\ (\ref{eq:Vdef3D}) and (\ref{eq:Cdef3D}), this gives a voltage
\be
V - V_t \simeq \frac{e}{4 \pi \varepsilon_0 \varepsilon R_s}  \exp \left[ - \frac{1}{N^{1/3} R_s} \right]
\label{eq:V-smallN-3D}
\ee
and a capacitance
\be
C(N) \simeq 12 \pi \varepsilon_0 \varepsilon \mathbb{V} N^{4/3} R_s^2 \exp \left[ \frac{1}{N^{1/3} R_s} \right].
\label{eq:C-smallN-3D}
\ee
Combining these two relations gives an expression for the volumetric capacitance as a function of voltage, applicable at very small $V - V_t$:
\be
\mathbb{C} (V) \simeq \frac{3}{R_s^3} \frac{e}{V - V_t} \ln^{-4} \left[ \frac{e/4 \pi \varepsilon_0 \varepsilon R_s}{V - V_t} \right].
\label{eq:C-smallV-3D}
\ee

Eq.\ (\ref{eq:C-smallV-3D}) implies that at small $V - V_t$, where ions are sparse, the capacitance can be much larger than the mean-field result of Eq.\ (\ref{eq:C3Dmf}).  This growth in the capacitance is driven by the vanishing interaction [Eq.\ (\ref{eq:Yukawa})] between discrete, correlated ions.  The maximum value of the capacitance occurs at $(V - V_t) \rightarrow 0$ and is determined by thermal effects.  This maximum can be estimated, as in the previous section, by making a virial expansion of the free energy
\be
F \simeq F_{id} + \mathbb{V} N^2 k_BT B(T) - Q (V -V_t).
\ee
The value of the capacitance in this limit, as in Eqs.\ (\ref{eq:Cvirialn})--(\ref{eq:Cmax}), is inversely related to the virial coefficient $B(T)$.  At not too small $R_s$, such that $R_s \gtrsim a/2$,
\be
\mathbb{C}_{max}(T) \simeq \frac{2 \pi \varepsilon_0 \varepsilon R_s}{\widetilde{T} B(T)}.
\label{eq:Cvirial3D}
\ee
Here, $\widetilde{T}$ is a dimensionless temperature normalized to the interaction between two charges at a distance $R_s$:
\be
\widetilde{T} = \frac{k_BT}{e^2/4\pi \varepsilon_0 \varepsilon R_s} = \frac{R_s}{d} T^* .
\ee
The virial coefficient $B(T)$ is calculated as
\begin{eqnarray}
B(T) &= & \frac12 \int_0^{\infty} \left(1 - \exp\left[-\frac{e \phi(r)}{k_BT}\right] \right) 4 \pi r^2 dr \\
& \simeq & 2 \pi R_s^3 \left( 1 + \ln[1/\widetilde{T}] + \frac13 \ln^3[1/\widetilde{T}] \right),
\label{eq:B3D}
\end{eqnarray}
so that Eq. (\ref{eq:Cvirial3D}) can be written
\be
\mathbb{C}_{max}(T) \approx \frac{1}{\widetilde{T} \left( 1 + \ln[1/\widetilde{T}] + \frac13 \ln^3[1/\widetilde{T}] \right)} \frac{\varepsilon_0 \varepsilon}{R_s^2}.
\label{eq:Cmax3D}
\ee

\begin{figure}[htb]
\centering
\includegraphics[width=0.45 \textwidth]{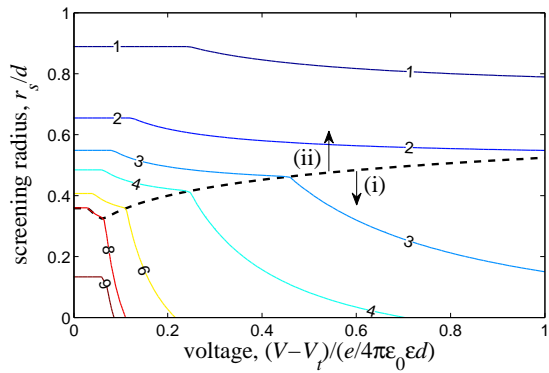}
\caption{(Color online).  A contour plot of the volumetric capacitance $\mathbb{C}$ as a function of voltage $V$ and electrode screening radius $r_s$ for charges in an electrode with planar pores of width $d$ separated by walls of thickness $b = d$ at room temperature.  Contours are labeled by their values of $\mathbb{C}$ in units of $\varepsilon_0 \varepsilon/d^2$.  The dashed line separates the regions of validity of the two theories presented in this section: (i) where pores are non-interacting and the pore width can be renormalized according to $d \rightarrow d_\text{eff} = d + 2 r_s$ (lower region), and (ii) where charges interact three dimensionally with the interaction law of Eq.\ (\ref{eq:Yukawa}) (upper region).  For $\varepsilon = 2$ and $d = 1$ nm, the unit of volumetric capacitance $\varepsilon_0 \varepsilon/d^2 = 18$ F/cm$^3$ and the unit of voltage $e/4 \pi \varepsilon_0 \varepsilon d = 0.7$ Volts.} \label{fig:volumeC-contours}
\end{figure}

If one wishes to formulate an approximate prediction for the capacitance at arbitrary values of screening radius, including values of $r_s$ that are comparable to $b$, then one may evaluate separately the capacitance based on approach (i), where the 2D pore thickness is renormalized as $d \rightarrow d_\text{eff}$, and approach (ii), where ions interact three-dimensionally, and then take the smaller value.  Since intra- and inter- pore interactions contribute additively to the total energy, these create series contributions to the capacitance, so that as a zero-order approximation one can take the smaller of the two capacitances.  The result of this process is shown in Fig.\ \ref{fig:volumeC-contours}, which constitutes a prediction for the volumetric capacitance $\mathbb{C}$ at arbitrary voltage and electrode screening radius.

\section{Nonlinear screening in graphite (carbon) supercapacitors} \label{sec:graphite}

The preceding sections outline a general theory for the volumetric capacitance of a supercapacitor made with electrodes that can either be considered metallic or can be characterized by some linear screening radius $r_s$.  In this section we discuss specifically the case of graphite electrodes, which is among the most commonly-studied materials for supercapacitor devices \cite{Chmiola2006aii, Largeot2008rbi, Wang2006eoh, Simon2008mec}.

In graphite, the Fermi level density of states $\nu(\mu_F)$ is actually relatively small, so that the linear screening radius $r_s \approx 0.8$ nm cannot be considered much smaller than the spacing between pores \cite{Gerischer1985iod}.  For example, if $d = b = 1$ nm, then Fig.\ \ref{fig:volumeC-contours} would seem to imply a capacitance on the order of $\varepsilon_0 \varepsilon/d^2 \approx 18$ F/cm$^3$.  Experiments with graphite electrodes, however, yield a capacitance five times larger than this value \cite{Simon2008mec, Largeot2008rbi}, suggesting that graphite screens over a much smaller distance than $r_s$ and effectively behaves as a good metal.

This apparent discrepancy can be resolved if one recalls that the density of states in graphite is close to $\nu(\mu_F)$ only in a narrow range of energies, beyond which it increases linearly with energy on both sides of the Fermi level \cite{Gerischer1985iod}.  Such variation of the density of states suggests that screening by graphite is \emph{non-linear} even when a relatively small electric field is applied to the surface of the pore, and that therefore the screening properties of the electrode material cannot be characterized by a constant linear screening radius $r_s$.

In order to estimate the distance over which the ions' potential is screened, one can consider the problem of a uniform applied electric field $\vec{E_0}$ orthogonal to the basal plane of graphite.  It has been shown \cite{Pietronero1978cdi} that in this case the magnitude of the electric field $E(z)$ decays with the distance $z$ beyond the graphite interface as
\be
E(z) = \frac{E_0}{(1 + z/2z_0)^3},
\ee
where 
\be
z_0 = \frac{\sqrt[3]{3}}{2} c \left( \frac{e}{4 \pi \varepsilon_0 \varepsilon \alpha^2 c^2 E_0} \right)^{1/3}
\label{eq:z0}
\ee
is the centroid location of the counter-charge in the graphite (the non-linear screening radius),  $c \approx 0.34$ nm is the distance between graphite planes (graphene sheets), and $\alpha \approx 2.2/\varepsilon$ is the effective fine structure constant of graphene.  (See also recent discussions of screening in graphene multilayers in Refs.\ \onlinecite{Guinea2007cda, Koshino2010ise}).

The implications of this result for the problem of screening of adsorbed ions in a graphite pore can be seen as follows.  When the area density of ions inside a pore is large enough that $na^2 \simeq 1$, these ions can be said to produce a roughly uniform electric field at the wall of the pore whose strength is $E_0 \simeq e/2 \varepsilon_0 \varepsilon a^2$.  Inserting this relation into Eq.\ (\ref{eq:z0}) yields a screening distance $z_0 \simeq 0.75 c$, which suggests that the field is entirely screened within the first graphene layer.  In other words, at dense ion filling the electric field between ions in the pore does not penetrate beyond the first graphene layer.  Therefore, despite its relatively low density of states $\nu(\mu_F)$, graphite may be treated as a metal at not-too-small ion densities $na^2 \gtrsim 0.5$ (note that $na^2 > 0.5$ occupies the majority of the voltage range in Fig.\ \ref{fig:C-V}). 

We can also discuss what happens with the volumetric capacitance when
the density of ions is smaller ($V - V_t$ decreases), so that the electric field $E_0$ produced by the ions becomes weaker.  When $V-V_t$ is made moderately small, the effective density of states decreases, the nonlinear screening radius $z_0$ grows, and the nonlinear capacitance decreases. One can show using  Eq.\ (\ref{eq:z0}) that the capacitance $C \propto (V-V_t)^{1/2}$. Eventually, at small $V-V_t$, the effective density of states saturates at the level of $\nu(\mu_F)$, so that the screening radius becomes constant and equal to the linear screening radius $r_s \approx 0.8$ nm.  In this limit, the volumetric capacitance is relatively small and is given by the theory of Sec.\ \ref{sec:3Dscreening}, as was already discussed in the beginning of this section.

\section{Capacitance of a crystalline assembly of metallic spheres} \label{sec:balls}

So far we have restricted our discussion to the electrode geometry shown in Fig.\ \ref{fig:comb}.  In practice, such electrodes with parallel planar pores are difficult to make. In many cases supercapacitor electrodes are simply a random assembly of conducting particles, arranged so that the particles form an infinite, conducting cluster through which electrons can percolate while the pores in this cluster form a separate percolating space through which the ionic liquid can freely pass.  For such cases the model of Fig.\ \ref{fig:comb} is a strong idealization. In this section we would like to briefly discuss another idealized electrode structure which captures the three-dimensional character of pores. 

Consider an assembly of metallic nanospheres, each with the same radius $R$, arranged so that they form a cubic lattice with nearest neighbor spheres touching each other.  As in previous sections, we imagine that this crystalline film is deposited onto a metallic contact plate, connected to a voltage source, and immersed in an ionic liquid.  A voltage $V$ is applied between the contact plate and the bulk of the ionic liquid.  If the diameter of the ions within the ionic liquid is small enough, then this arrangement produces an effective supercapacitor electrode, where ions may percolate through the spaces between conducting spheres and neutralize the electronic charge provided by the voltage source.

In order to calculate the capacitance of this electrode, we first analyze the interaction between two ions that enter into the bulk of the electrode.  This can be done by calculating the potential as a function of distance produced by a single ion in the center of a pore deep inside the electrode bulk.  We calculate this potential numerically using the relaxation method for solving the Laplace equation, where each of the conducting spheres is held at zero potential.  We find that potential decays exponentially with radial distance from the ion with a characteristic screeening length $R_s = R \times (0.26 \pm 0.01)$.  The reason for this sharp decay is the same as for the decay of the potential in 2D, slit-like pores [Eq.\ (\ref{eq:phi})]: electric field lines emanating from the ion are adsorbed by the surface of nearby conducting spheres, and the number of these field lines that survive by passing through the narrow, tortuous pores between spheres decays exponentially with distance. 

Once the interaction law is known, one can calculate the capacitance in a way similar to the analysis of Sec.\ \ref{sec:3Dscreening}.  We arrive then at a relation $C(V)$ which has the maximum given by Eq.\ (\ref{eq:Cmax3D}).  For $R=4$ nm, $T=300$ K, and $\varepsilon = 2 $ this relation gives a volumetric capacitance $\mathbb{C}_{max} = 1.7 \varepsilon_0\varepsilon/R_s^{2} = 25 \varepsilon_0\varepsilon/R^2 \approx 28$ F/cm$^3$.  Remarkably, in this arrangement the capacitance per sphere is roughly 16 times larger than the capacitance of a single, isolated sphere in a medium with dielectric constant $\varepsilon$.  

One can reach even larger volumetric capacitance if the spheres are densely packed rather than arranged in a cubic lattice.  Reducing the radius $R$ of the spheres also sharply increases the capacitance.  In our next publication we will explore these mechanisms for increasing the capacitance by combining MC modeling with the analytical estimates presented in Sec.\ \ref{sec:3Dscreening}.

\vspace*{2ex} \par \noindent
{\em Acknowledgments.}

We are grateful to M.M.\ Fogler, Yu.\ Gogotsi, A.\ Kamenev, and A.\ Stein for helpful discussions.  B.S. acknowledges the support of the NSF and M.S.L. thanks the FTPI for financial support.

\bibliography{nanopores}

\end{document}